\documentclass[preprint2]{emulateapj}

\usepackage{graphicx,amsmath,txfonts,url}
\usepackage{bm}
\graphicspath{{./fig/}{./png/}}

\newcommand{\EQ}{\begin{equation}}
\newcommand{\EN}{\end{equation}}
\newcommand{\EQA}{\begin{eqnarray}}
\newcommand{\ENA}{\end{eqnarray}}
\newcommand{\eq}[1]{(\ref{#1})}

\newcommand{\Eq}[1]{Equation~(\ref{#1})}
\newcommand{\Eqs}[2]{Equations~(\ref{#1}) and~(\ref{#2})}

\newcommand{\App}[1]{Appendix~\ref{#1}}
\newcommand{\Sec}[1]{Section~\ref{#1}}

\newcommand{\Fig}[1]{Figure~\ref{#1}}

\newcommand{\Figs}[2]{Figures~\ref{#1} and \ref{#2}}

\newcommand{\Tab}[1]{Table~\ref{#1}}

\newcommand{\bra}[1]{\langle #1\rangle}

{}
{}
{}

{}
{}
{}
{}
{}
{}
{}
{}
{}
\newcommand{\meanBB}{\overline{\mbox{\boldmath $B$}}{}}{}
{}
{}
{}
{}
{}
{}
{}
{}
{}
{}

{}
{}
{}

\newcommand{\meanB}{\overline{B}}

\newcommand{\meanF}{\overline{F}}

{}

{}
{}

\newcommand{\xxx}{\hat{\mbox{\boldmath $x$}} {}}
\newcommand{\yyy}{\hat{\mbox{\boldmath $y$}} {}}

\newcommand{\xx}{\bm{x}}

\newcommand{\uu}{\mbox{\boldmath $u$} {}}
\newcommand{\UU}{\mbox{\boldmath $U$} {}}

\def\bb{\bm{b}}
\newcommand{\BB}{\mbox{\boldmath $B$} {}}
\newcommand{\EE}{\mbox{\boldmath $E$} {}}
\newcommand{\jj}{\mbox{\boldmath $j$} {}}
\newcommand{\JJ}{\mbox{\boldmath $J$} {}}

\newcommand{\AAA}{\mbox{\boldmath $A$} {}}

\newcommand{\ee}{\mbox{\boldmath $e$} {}}

\newcommand{\ff}{\mbox{\boldmath $f$} {}}

\newcommand{\FF}{\mbox{\boldmath $F$} {}}

\newcommand{\nab}{\mbox{\boldmath $\nabla$} {}}

\newcommand{\SSSS}{\mbox{\boldmath ${\sf S}$} {}}

\newcommand{\EMF}{\mbox{\boldmath ${\cal E}$} {}}

\newcommand{\DDD}{{\cal D} {}}

\newcommand{\const}{{\rm const}  {}}

\def\Sh{\mbox{\rm Sh}}

\def\Pm{\mbox{\rm Pr}_M}
\def\Rm{\mbox{\rm Re}_M}

\def\Rey{\mbox{\rm Re}}

\def\cs{c_{\rm s}}

\def\hf{h_{\rm f}}

\def\kf{k_{\rm f}}

\def\urms{u_{\rm rms}}

\def\etat{\eta_{\rm t}}
\def\etatz{\eta_{\rm t0}}

\def\Beq{B_{\rm eq}}

\def\half{{\textstyle{1\over2}}}

\def\onethird{{\textstyle{1\over3}}}

\newcommand{\yapj}[3]{ #1, {ApJ,} {#2}, #3}

\newcommand{\yana}[3]{ #1, {A\&A,} {#2}, #3}

\newcommand{\ymn}[3]{ #1, {MNRAS,} {#2}, #3}

\newcommand{\ypre}[3]{ #1, {Phys.\ Rev.\ E,} {#2}, #3}

\newcommand{\yjour}[4]{ #1, {#2}, {#3}, #4}

\newcommand{\ybook}[3]{ #1, {#2} (#3)}

\newcommand{\smn}[1]{ #1, {MNRAS}, submitted}
\newcommand{\parr}[2]{\frac{\partial #1}{\partial #2}}
\newcommand{\hhh}{h}
\newcommand{\hhM}{\overline{h}}
\newcommand{\FFM}{\overline{\bm{F}}}
\newcommand{\FFp}{\FF_{\Phi}}
\newcommand{\FFad}{\FF_{\text{adv}}}
\newcommand{\FFr}{\FF_{\text{res}}}
\newcommand{\FFd}{\FF_{\text{dyn}}}
\newcommand{\FFf}{\FF_{\text{f}}}

\newcommand{\FFfM}{\FFM_{\text{f}}}
\newcommand{\FFmM}{\FFM_{\text{m}}}

\newcommand {\AAAM}{\overline{\AAA}}
\newcommand {\BBM}{\overline{\BB}}
\newcommand {\JJM}{\overline{\JJ}}
\newcommand {\UUM}{\overline{\UU}}
\newcommand {\EMFM}{\overline{\EMF}}
\newcommand {\EEM}{\overline{\EE}}
\newcommand{\aaa}{\mbox{\boldmath $a$}{}}{}
\newcommand{\PHIM}{\overline{\Phi}}
\newcommand{\hhm}{\overline{h}_{\text{m}}}
\newcommand{\hhf}{\overline{h}_{\text{f}}}
\newcommand{\FVC}{\FFM_{\text{VC}}}
\newcommand{\CVC}{C_{\text{VC}}}
\newcommand{\kappaa}{\kappa_\alpha}

\begin{document}

\title{Magnetic helicity flux in the presence of shear}
\author{Alexander Hubbard$^1$ and Axel Brandenburg$^{1,2}$}

\email{alex.i.hubbard@gmail.com
($ $Revision: 1.81 $ $)
}

\affil{
$^1$ NORDITA, AlbaNova University Center, Roslagstullsbacken 23,
SE 10691 Stockholm, Sweden\\
$^2$Department of Astronomy, AlbaNova University Center,
Stockholm University, SE 10691 Stockholm, Sweden
}

\begin{abstract}
Magnetic helicity has risen to be a major player in dynamo theory, with the helicity
of the small-scale field being linked to the dynamo saturation process
for the large-scale field.
It is a nearly conserved quantity, which allows its evolution equation
to be written in terms of production and flux terms.
The flux term can be decomposed in a variety of fashions.
One particular contribution that has been expected to play a significant
role in dynamos in the presence of mean shear was isolated by
Vishniac \& Cho (2001, ApJ 550, 752).
Magnetic helicity fluxes are explicitly gauge dependent however, and the correlations
that have come to be called the Vishniac--Cho flux were determined in the
Coulomb gauge, which turns out to be fraught with complications
in shearing systems.  While the fluxes of small-scale helicity are explicitly gauge dependent, their
divergences can be gauge independent.
We use this property to investigate magnetic helicity fluxes of small-scale
field through direct numerical simulations in a shearing-box system and find
that in a numerically usable gauge the divergence of the small-scale helicity flux vanishes, while
the divergence of the Vishniac--Cho flux remains finite.
We attribute this seeming contradiction to the existence of horizontal
fluxes of small-scale magnetic helicity with finite divergences.
\end{abstract}


\section{Introduction}

The large-scale magnetic field of the Sun and other stars is often
modeled using mean-field theory \citep{RH04}.
Important ingredients in this theory are the $\alpha$ effect responsible
for field amplification and an enhanced (turbulent) magnetic diffusivity
\cite{Mof78,KR80}.
Once the field has reached appreciable field strength, these effects
become modified through the backreaction of the Lorentz force.
Often a simple algebraic quenching formula is being assumed, but such
a simple prescription is unable to model correctly the quenching under
more general conditions with shear \citep{BBS01} or boundaries \citep{BD01}.

Over the past decade, significant progress has been made in modeling the dynamo saturation process
in mean-field models through the development and use of 
the dynamical $\alpha$ quenching methodology.
This methodology \citep[originally due to][]{KR82} has explained
several puzzling features of MHD dynamos, such as the slow saturation phase of a homogeneous
$\alpha^2$ dynamo \citep{FB02,BB02}, and has reposed the crucial question
of catastrophic $\alpha$ quenching \cite[see][for a review]{BS05}.
In this picture, the saturation of the dynamo is caused
by the build-up of magnetic helicity, which is
nearly conserved in the high conductivity limit in the absence of fluxes.  This raises the possibility of
speeding up the saturation process and reaching significant saturation field strength
through mechanisms that export or destroy magnetic helicity.

Making general use of the dynamical $\alpha$ quenching methodology in open systems
then requires an understanding of magnetic helicity fluxes, more significantly
an understanding of the fluxes of magnetic helicity of the small-scale field.
In the following we often refer to such fluxes as small-scale magnetic
helicity fluxes, although this is not quite accurate,
because it is itself a mean quantity and not a fluctuation.
Recent work has shown that in inhomogeneous systems there is a turbulent
diffusive flux of small-scale helicity, at least in the absence
of shear, although the diffusion coefficient is in some cases smaller than
expected \citep{Mitra10,HB10}.
Consideration of that flux term has allowed mean-field models to capture
the saturation
behavior of some non-triply periodic, non-homogeneous dynamos without shear.  Unfortunately, the fluxes of small-scale
magnetic helicity are explicitly gauge dependent, and in the presence of turbulence, can be decomposed
in different fashions.

In \citet{VC01}, an interesting component of the flux of small-scale helicity was isolated.  This component
has been named the Vishniac--Cho flux, henceforth the VC flux.
In later work \citep{SB04,BS05a}, the
form of this flux in the presence of uniform shear was calculated, and found to be both simple and of significant
magnitude.  Shear drives an $\Omega$ effect and is an important and nearly omni-present player in astrophysical
dynamos, so the VC flux has seen significant interest,
both in mean-field modeling \citep{BS05a} and in the interpretation of
the differences between similar direct numerical simulations with
differing boundary conditions \citep{B05,KKB08}.
It is important to realize however that
shear poses unique difficulties in the formulation, and importantly, interpretation, of magnetic helicity fluxes.
It is the goal of this paper to explore those difficulties and determine
the significance of the VC flux.
We use the perhaps surprising result that while magnetic helicity fluxes are gauge dependent, their divergences
may not be across broad gauge-families \citep{Mitra10} to allow us to compare the VC flux
with the small-scale magnetic helicity flux in a gauge where the mean shear is easy to treat.  Our investigations
will bear weight on the interpretation and use of the VC flux, but we will not and indeed cannot extract the VC
flux from the simulations we perform.

In \Sec{maghelflux} we discuss the various contributions to the
magnetic helicity flux and define our mean-field decomposition.
Further, we explain the broad gauge independence
of small-scale magnetic helicity flux divergences.
In \Sec{shear} we sketch the difficulties inherent in uniform shear, define the shearing-advective gauge
and derive the small-scale magnetic helicity flux in that gauge. 
In \Sec{VC} we discuss the VC flux as calculated in \cite{SB04} and \cite{BS05a}.
In \Sec{Sim} we present the results of our direct numerical simulations and compare the results with
the VC flux.
We discuss the significance of our results in \Sec{comparison} and conclude in \Sec{conclusion}.

\section{Magnetic helicity flux and formalism}
\label{maghelflux}

We begin by deriving the formula for magnetic helicity fluxes in general.  The MHD equations
for the magnetic field are:
\begin{eqnarray}
&&\BB=\nab \times \AAA, \label{Adef}\\
&&\JJ=\nab \times \BB/\mu_0, \label{J} \\
&&\EE=-\UU\times \BB+\eta \mu_0\JJ \label{electric}, \\
&&\parr{\AAA}{t}=-\EE-\nab\Phi \label{dAdt},
\end{eqnarray}
where $\eta$ is the molecular resistivity, $\mu_0$ is the vacuum
permeability, and $\Phi$ is the electrostatic or scalar potential
that determines our gauge.
For example, setting $\Phi=0$ results in the
Weyl gauge, of interest numerically because it simplifies \Eq{dAdt},
while a solution of \Eq{dAdt} with
$\nab^2 \Phi=-\nab \cdot \EE$ will have constant $\nab \cdot \AAA$
and with an appropriate initial condition on $\AAA$ results
in the Coulomb gauge.

The time evolution of the magnetic helicity density
$\hhh \equiv \AAA \cdot \BB$ is then given by
\begin{eqnarray}
\parr{\hhh}{t}&=&\parr{\AAA}{t}\cdot\BB+\AAA\cdot\parr{\BB}{t} \nonumber \\
&=&-2 \eta \mu_0\JJ \cdot \BB -\nab \cdot \left(\EE \times \AAA+\Phi \BB \right). \label{dHdt}
\end{eqnarray}
The flux $\FFp$ of magnetic helicity in a given gauge with a corresponding
$\Phi$ can be read out of \Eq{dHdt}:
\begin{eqnarray}
\FFp&=&\EE \times \AAA +\Phi \BB=-\left(\UU \times \BB\right)\times \AAA+\Phi \BB +\eta \mu_0\JJ \times \AAA 
\nonumber \\
&=&\hhh \UU +(\Phi-\UU \cdot \AAA)\BB +\eta \mu_0\JJ \times \AAA.
\label{Fp}
\end{eqnarray}
In \Eq{Fp} we recognize the advective flux $\FFad \equiv \hhh \UU$,
a resistive flux $\FFr \equiv \eta \mu_0\JJ \times \AAA$, and finally
a dynamical flux,
\EQ
\FFd \equiv (\Phi-\UU\cdot\AAA)\BB.
\label{FFd}
\EN
The formula for $\FFd$
leads us to consider  ``advective'' gauges of the form $\Phi\equiv\UU' \cdot \AAA$, where $\UU'$ is one
component of the velocity field, see \Sec{advgauge}.

\subsection{Mean-field decomposition}

We proceed to a mean-field decomposition of the magnetic helicity flux.
We denote general averaging schemes by overbars.
Fluctuating terms will be denoted by lower cases or primes:
\EQ
\AAA=\AAAM+\aaa,\quad
\AAA \cdot \BB=\overline{\AAA \cdot \BB}+\left(\AAA \cdot \BB\right)'.
\EN
The mean-field decomposition of \Eqs{dAdt}{dHdt} yields
\EQ
\parr{\AAAM}{t}=\UUM \times \BBM +\EMFM-\eta \mu_0\JJM -\nab \PHIM,
\EN
where $\EMFM \equiv \overline{\uu \times \bb}$ and
\EQ
\parr{\hhM}{t}=-2\eta\mu_0\overline{\JJ \cdot \BB}-\nab \cdot \overline{\EE \times \AAA}
-\nab \cdot \overline{\Phi \BB}.
\EN
The latter can be written as
\EQ
\parr{\hhM}{t}=\parr{\hhm}{t}+\parr{\hhf}{t},
\EN
where $\hhM=\hhm+\hhf$, with $\hhm \equiv \AAAM \cdot \BBM$ being the
helicity in the large-scale fields and $\hhf \equiv \overline{\aaa \cdot \bb}$
the helicity in the small-scale fields.

The evolution equations for these helicities are
\begin{eqnarray}
\parr{\hhm}{t}=&+&2\EMFM \cdot \BBM -2\eta \mu_0\JJM \cdot \BBM -\nab \cdot \left(\EEM \times \AAAM+\PHIM \BBM\right),
\label{dhmdt}\\
\parr{\hhf}{t}=&-&2 \EMFM \cdot \BBM-2\eta \mu_0\overline{\jj \cdot \bb}
-\nab \cdot\left(\overline{\ee \times \aaa}+\overline{\phi \bb}\right),
\label{dhfdt}
\end{eqnarray}
where $\phi=\Phi-\PHIM$ is the fluctuating scalar potential and
\EQ
\EMFM=\alpha \BBM-\etat \mu_0 \JJM
\label{EMFM}
\EN
is the mean turbulent electromotive force, where $\alpha$ is the
$\alpha$ effect and $\etat$ is the turbulent magnetic diffusivity.
From \Eqs{dhmdt}{dhfdt} we find the fluxes of the large-scale and
small-scale fields:
\begin{eqnarray}
\FFmM &=& \EEM \times \AAAM+\PHIM \BBM \label{FFm}, \\
\FFfM &=& \overline{\ee \times \aaa}+\overline{\phi \bb} \label{FFfeq}.
\end{eqnarray}

\subsection{The significance of gauges for magnetic helicity fluxes}
\label{significance} 
 
It is clear from the form of $\FFd$ in \Eq{FFd} that any consideration of helicity
fluxes must also take into account the gauge choice, as at any point
where $\FFd \neq 0$, the value of $\FFd$ can be arbitrarily set by the
gauge or, equivalently, the condition for $\FFd$ being independent of
$\Phi$ is that $\FFd=0$ for all gauges.
A gauge choice that generates a desired flux along $\bm{\hat{x}}$
is always possible provided $\BB \cdot \hat{\bm{x}} \neq 0$.
Such a gauge takes the form $\Phi(\xx,t)=f(\xx,t) B_x$.
Recall also that boundary or
symmetry conditions on the physical system do not apply to the vector potential or the gauge (although
numerical simulations may require gauge choices where they do).  The ability to add an arbitrary
flux of magnetic helicity to the system via the addition of a new gauge makes the isolation of differing components
of the flux a risky business.
 
There are effects that mitigate this gauge dependence however.
The divergence of $\FFf$ and the term $\partial\hhM/\partial t$ are the
only gauge-dependent terms in \Eq{dhfdt}.
If $\hhf$ is indeed gauge
independent then the \emph{divergence} of $\FFf$ must be gauge independent as well,
even though the flux itself is explicit gauge dependent.
Clearly the divergence of $\FFf$ is the same for all gauges for which
$\partial_t \hhf$ is the same.
As long as our shearing-box has a time-constant $\hhf$ in the saturated regime,
we can make statements about the divergence of $\FFf$ for all gauges which would have a time-constant
saturated $\hhf$.

The dynamical $\alpha$ quenching methodology, one of the primary consumers of magnetic helicity
information, assumes that the small-scale magnetic helicity $\hhf$ and
the small-scale current helicity $\overline{\jj \cdot\bb}$ are proportional.
This requirement is often used as an argument in favor of the Coulomb
gauge \citep{KR99}.
Even if the saturated current helicity is not time-independent (as might be the case for an oscillating solution),
as the current helicity is gauge independent,
the validity of the dynamic $\alpha$ quenching methodology assumes therefore that the former is as well.
Recent studies have supported this hypothesis \citep{Mitra10,HB10},
at least in the limits of numerical simulations, which disallow extreme levels of gauge pathology by forcing
the vector potential to be numerically resolved.
(This will be discussed in more detail in a separate paper where
we solve an evolution equation for the gauge transformation.)
Alternatively, one could restrict oneself to families of gauges
where the relation holds.  We note that, in that regard, both the magnetic and current helicities are shearing-periodic
in our system in the shearing-advective gauge that will be described later,
while in the Weyl gauge the magnetic helicity is not . 

\section{Shear}
\label{shear}

The presence of shear poses further difficulties when considering magnetic helicity fluxes.  To see this, consider
a shearing periodic box with an imposed flow, $\UU_S=(0,Sx,0)$,
and sides of length $L$ centered
on the origin.  We will use the Weyl gauge ($\Phi=0)$
and a planar averaging scheme:
\EQ
\BBM(x,y,z,t) \equiv L^{-2}\int_{-L/2}^{L/2} \int_{-L/2}^{L/2} dx' dy' \BB(x',y',z,t).
\EN
In what follows, we will assume
that the shear flow is the only large-scale velocity, and note that it averages to $0$ and so technically
is a fluctuating field.
This could in principle be avoided under a local planar average over
a square centered on $(x,y)$ for example \citep{BRRK08}.
However, such an average is problematic too, because it does not obey
one of the Reynolds rules: the average of a product of an average and a
fluctuation does not vanish.
Nevertheless, even though uniform shear can complicate averaging
schemes, it is easier to treat than non-uniform shear with the resulting
non-uniform $\Omega$ effect.
We therefore proceed with our standard (non-sliding) averaging scheme.
We will also assume
that, at the single instant in time that we consider, the vector potential $\AAA$ is shearing periodic as well.  Note that
in the Weyl gauge, the vector potential will not remain shearing periodic.
Numerical simulations in the shearing box approximation use therefore a
different gauge \citep{BNST}, as will be discussed below.

The difficulty in treating the helicity flux can be seen from \Eq{Fp} which,
in the Weyl gauge, becomes
\EQ
\FF= \hhh \UU_S -( \UU_S \cdot \AAA) \BB
+\hhh \uu -(\uu \cdot \AAA) \BB+\eta \mu_0\JJ \times \AAA.
\label{FSS1}
\EN
If $\AAA$ were shearing periodic, the last four terms of \Eq{FSS1}
would be likewise shearing periodic
and hence would not contribute a net divergence to the system.
However, the first two terms on the right-hand side of \Eq{FSS1}
violate shearing-periodicity.
In particular, the $x$ component of the second term on the right-hand side
of \Eq{FSS1} reduces to
\EQ
F_x=-( \UU_S \cdot \AAA) B_x+...=-Sx A_y B_x +...,
\label{shearhorflux}
\EN
where the dropped terms cannot contribute a net divergence.
Recall that, at first, $A_y B_x$ would here still be shearing periodic.
Systems with shearing-periodic magnetic vector potentials would then allow
for a \emph{horizontal} flux of magnetic helicity.

\subsection{Helicity fluxes in advective gauges}
\label{advgauge}

To examine the VC flux numerically, we adopt a homogeneous
shearing-periodic setup.
(The resulting magnetic field will however become inhomogeneous and could
produce finite magnetic helicity fluxes and flux divergences.)
As discussed above,
to keep the magnetic vector potential itself shearing-periodic we must use an appropriate gauge, namely
$\Phi=\UU_S \cdot \AAA$, which we term ``shearing-advective''.
This is also the gauge used by \cite{BNST}.
More generally, we can define a family of ``advective gauges'' with $\Phi_A=\UU_A \cdot \AAA$,
for a component of the velocity $\UU_A$, with the corresponding $\UU_{NA}=\UU-\UU_A$.
The name ``advective'' is chosen because in this gauge the effect of $\UU_A$ on the helicity
flux is advective as can be seen from \Eq{Fp}, which becomes:
\begin{eqnarray}
\EE_A&=&-\UU_{NA}\times \BB+\eta \mu_0\JJ, \\
\FFp&=&h\UU_A+(\EE_A \times \AAA).
\end{eqnarray}
If $\UU_A$ is a mean flow ($\overline{\UU}_A=\UU_A$), then the mean flux of the small-scale helicity
becomes
\EQ
\FFfM=\hhf \UUM_A+\overline{\ee_A \times \aaa}.
\label{FFfeq2}
\EN
Alternatively, if $\UU_A$ is not a mean flow ($\overline{\UU}_A=0$),
then we have
\EQ
\FFfM=\overline{\hf'\UU_A'}+\overline{\ee_A \times \aaa}.
\label{FFfeq2b}
\EN
For our system, with $\UU_A=\UU_S$, $\ee_A$, $\aaa$ and $\bb$ are all shearing-periodic,
and so their mean values, as well as all other mean values, are functions
of $z$ alone.  Correspondingly, only the $z$ component of $\FFf$ can have a finite divergence, and
we have eliminated the worry of horizontal magnetic helicity fluxes.  Further, to the extent that the system is
homogeneous, and invariant under a $180$ degree rotation about the $z$ axis, the horizontal fluxes vanish entirely
except for the advective flux due to the shear flow.

\section{The Vishniac--Cho flux with mean shear}
\label{VC}

The VC flux, $\FVC$, has been calculated in several places using
the first order smoothing approximation and later the $\tau$ approximation.
Their applicability to highly turbulent systems with large magnetic
Reynolds numbers cannot be guaranteed and is subject to verification
by numerical simulations, although it should work in cases
of small magnetic Reynolds numbers considered in this paper.
This flux was originally calculated in the Coulomb gauge, but it
can be calculated in a related gauge in which the magnetic helicity density
corresponds to a density of magnetic linkages \citep{SB06}.
It is
most interesting in the case of shear, and we will restrict ourselves to the consideration of
uniform shear which, as noted above, raises concerns about horizontal fluxes
with finite divergence.
In this system, $\FVC$ was calculated in the appendix of \citet{BS05a} to be
\EQ
\FVC=\CVC \,{S\hat{\bm{z}}\over2\kf^2}\, (\meanB_x^2-\meanB_y^2),
\label{FVC}
\EN
where $\CVC$ is a coefficient expected to be of order unity
and $\kf$ is the wavenumber of the energy-carrying eddies.
As eluded to in \Sec{significance}, we assume here that the current
helicity is proportional to $\kf^2$ times the magnetic helicity density.
There are other components of the magnetic helicity flux known, 
for example the term in \Eq{shearhorflux} can be
found in the $\FF^{\rm bulk}$ of \citet{SB06}, and, as noted in
\Sec{significance},
the interesting value is the divergence of the total of all fluxes,
which might pick up contributions from the $x$ component of the flux.

While the VC flux is of pressing interest to mean-field dynamo theory
and it has been invoked in the interpretation of numerical simulations
\citep{KKB08}, the work on its implications in mean-field theory is not
well developed.
In \Sec{MeanField} we present a brief mean-field
analysis of the effects of a VC flux in our
shearing-sheet system using \eq{FVC},
and compare those results to the work of \citet{GCB10} in spherical shells. 

\section{Model calculations}
\label{Sim}

\subsection{Preliminary considerations}
\label{travelingwave}

In order to quantify shear-driven magnetic helicity fluxes,
we consider the shearing box approximation \citep{WT88} with
periodic boundary conditions in the $y$ and $z$ directions and
shearing-periodic boundary conditions in the $x$ direction.
According to \Eq{FVC} we expect a magnetic helicity flux in the
$z$ direction.
However, because our system is periodic in the $z$ direction,
there will be no net magnetic helicity flux in or out of the domain.
Nevertheless, the local divergence of $\FVC$ should be finite
because, contrary to homogeneous $\alpha^2$ dynamos without shear,
$\meanB_x^2-\meanB_y^2$ is in general $z$-dependent for $\alpha\Omega$ dynamos.
Indeed, the mean field that develops is a reasonable
approximation to an $\alpha \Omega$ dynamo, where $|S|>|\alpha k_z|$.
The marginally excited kinematic solution of a mean-field
$\alpha \Omega$ dynamo is a traveling wave \citep[see, e.g.,][]{BS02}, i.e.\
\begin{equation}
\BBM=B_0\left(\sin \theta, \sqrt{2} \left|\frac {c}{\alpha}\right|
\sin(\theta+\chi), 0\right)
\end{equation}
with
\begin{equation}
c=\pm\left|\frac{\alpha S}{2 k_z}\right|^{1/2}\!\!=\pm\eta_T k_z, \quad
\theta=k_z(z-ct),\quad \chi=\mp{3\over4}\pi,
\label{wavespeed}
\end{equation}
where the sign in front of $c$ is given by the sign of the product $\alpha S$
and $B_0$ is an undetermined amplitude factor.
Note that the magnetic helicity of this large-scale field,
$\hhm=|c/\alpha| k_z^{-1} B_0^2$, is independent
of $z$ for the ``natural'' vector potential
\EQ
\AAAM=k_z^{-1} B_0 \left(-\sqrt{2} \left|\frac {c}{\alpha}\right|
\cos(\theta+\chi), \cos \theta,  0\right).
\EN
The point of this discussion is to emphasize that even for an initially
homogeneous system, \Eq{VC} would predict the appearance of a magnetic
helicity flux.
This flux would lead to the annihilation of magnetic helicity fluctuations
of opposite sign --- even if such fluctuations were not present initially.
The effect of such fluxes can be predicted by mean-field models with
catastrophic quenching included \citep{BCC09}.
As we will show in the next section, the VC flux has actually an adverse effect
on the saturation behavior, and that only fluxes with the opposite sign
are able to accelerate the saturation of the mean field.

\subsection{Mean-field model with diffusive and VC fluxes}
\label{MeanField}

To demonstrate the difference between
mean-field predictions with and without the presence of a VC flux that is not
compensated for by other fluxes we use a mean-field dynamical
$\alpha$ quenching model \citep{KR82,BB02,BCC09}.
This methodology combines the dynamical $\alpha$ quenching equations
\begin{equation}
\alpha(z,t)=\alpha_K+\alpha_M,\quad 
\alpha_M= \etat \kf^2 \frac{\hhf}{\Beq^2},
\end{equation}
\begin{equation}
\frac{\partial \alpha_M}{\partial t}=
-2\etat\kf^2\left(\frac{\EMFM \cdot \BBM}{\Beq^2}+\frac{\alpha_M}{\Rm}\right)-
\nab\cdot\overline{\FF}_{\alpha},
\end{equation}
with the standard mean-field equation (in the shearing-advective gauge with
$\UUM=0$):
\EQ
\frac{\partial \AAAM}{\partial t}=-S\overline{A}_y\xxx+\EMFM-\eta \mu_0\JJM,
\EN
where the $\EMFM$ is given by \Eq{EMFM}.

We test diffusive and VC fluxes, setting
\EQ
\overline{\FF}_{\alpha}=\frac{\etat \kf^2}{\Beq^2} \left(\FVC-\kappaa \nab \hhf \right).
\EN
We measure the strength of the kinetic $\alpha$ effect $\alpha_K$ and
the shear $S$ with the dynamo numbers
\EQ
C_{\alpha} \equiv \frac{\alpha_K}{\etat k_1} \simeq \frac{\kf}{k_1},
\quad C_S\equiv \frac{S}{\etat k_1^2},
\label{dyncoef}
\EN
where $k_1=2\pi/L_z$ is the minimal wavenumber of the domain
in the $z$ direction.
Note that $\alpha_K$ is assumed independent of $\meanBB$, so the kinetic
$\alpha$ effect is therefore also a kinematic one.

\begin{figure}[t!]\begin{center}
\includegraphics[width=\linewidth]{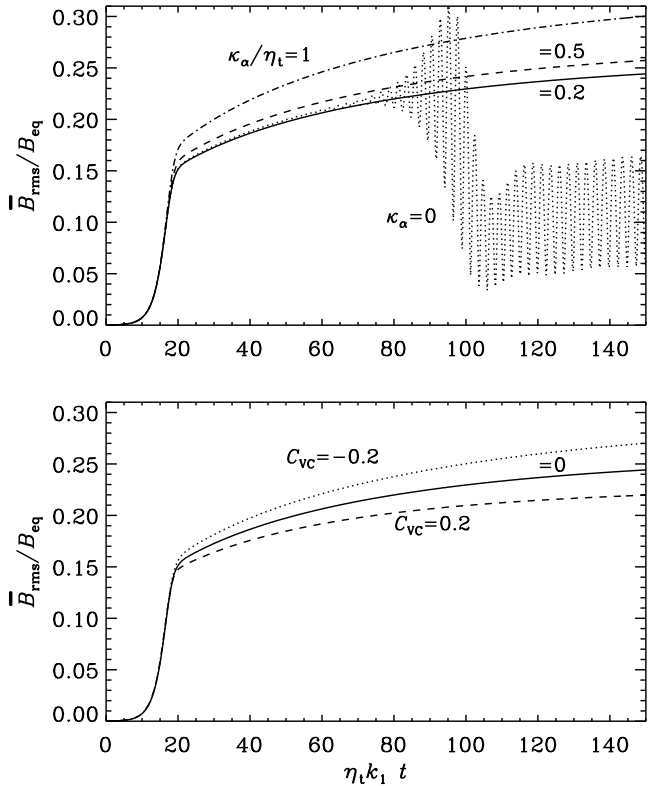}
\end{center}\caption{
Left:
Saturation behavior of models with $\etat/\eta=10^3$, $\CVC=0$,
$C_\alpha=-0.2$, and $C_S=-20$, and for $\kappaa / \etat$ ranging from 0 to 1.
Right:
Same, but for $C_{\rm VC}$ ranging from $-0.2$ to $+0.2$ and
$\kappaa / \etat=0.2$.
}\label{pcomp}\end{figure}

The results are shown in \Fig{pcomp}, where we plot the saturation
behavior of models with $\etat/\eta=10^3$, and dynamo numbers
$C_\alpha=-0.2$ and $C_S=-20$.
The left panel covers four values of $\kappaa/\etat$
 the turbulent diffusion coefficient for
$\hhf$, with the VC flux turned off ($\CVC=0$).  The right panel
displays the dynamo behavior for
three values of $C_{\rm VC}$ with $\kappaa/\etat=0.2$.
Note that in all calculations an early intermediate saturation level
of $\approx0.15\Beq$ is reached.
This is followed by a resistively slow saturation phase, as
was expected from models without shear \citep{B01,BB02}.
In agreement with earlier work, the saturation behavior is accelerated
by diffusive fluxes \citep{BCC09}.
In the absence of a diffusive flux, $\kappaa=0$, the field drops suddenly
back to lower values and continues to oscillate.
These oscillations are eliminated by small values of $\kappaa$ while not
significantly affecting the intermediate saturation behavior if
$\kappaa=0.2\etat$.
Note that for positive values of $\CVC$, the VC flux actually has
an adverse effect on the saturation behavior and only negative
values are able to accelerate the saturation.
This is similar to results for mean-field dynamo action in
spherical shells \citep{GCB10}.

\subsection{Simulations of shear flow turbulence}

We turn to the computation of magnetic helicity fluxes
through direct numerical simulations.
We solve the stochastically forced isothermal hydromagnetic equations
in a generally cubical domain of size $(2\pi)^3$ in the
presence of a uniform shear flow, $\UU_S=(0,Sx,0)$, with $S=\const$,
\EQ
{\DDD\AAA\over\DDD t}=-S A_y\xxx+\UU\times\BB+\eta\nabla^2\AAA,
\EN
\EQ
{\DDD\UU\over\DDD t}=-S U_x\yyy-c_{\rm s}^2\nab\ln\rho
+{1\over\rho}\JJ\times\BB+\FF_{\rm visc}+\ff,
\EN
\EQ
{\DDD\ln\rho\over\DDD t}=-\nab\cdot\UU,
\EN
where $\DDD/\DDD t=\partial/\partial t+(\UU+\UU_S)\cdot\nab$ is the
advective derivative with respect to the total flow velocity that
also includes the shear flow,
$\cs=\const$ is the isothermal sound speed,
$\FF_{\rm visc}=\rho^{-1}\nab\cdot(2\rho\nu\SSSS)$ is the viscous force,
${\mathsf S}_{ij}=\half(U_{i,j}+U_{j,i})-\onethird\delta_{ij}\nab\cdot\UU$
is the traceless rate-of-strain tensor, commas denote partial
differentiation, and $\ff$ is the forcing term.
As in earlier work \citep{B01} the forcing function consists of
plane polarized waves whose direction and phase change randomly from one
time step to the next.
The modulus of its wavevectors is taken from a band of wavenumbers
around a given average wavenumber that is referred to as $k_{\rm f}$.

The main control parameters in our simulations are the magnetic
Reynolds and Prandtl numbers, as well as the shear parameter,
\EQ
\Rm={u_{\rm rms}\over\eta k_{\rm f}},\quad
\Pm={\nu\over\eta},\quad
\Sh={S\over u_{\rm rms}\kf}.
\EN
We adopt periodic boundary conditions in the $y$ and $z$ directions
and shearing-periodic boundary conditions in the $x$ direction.
Our initial velocity, in addition to $\UU_S$, is $\UU=\bm{0}$ and the initial
density is $\rho=\rho_0\equiv\const$, while for the magnetic field we take
a Beltrami field of negative magnetic helicity and low amplitude
($10^{-7}$ times the equipartition value).
The magnetic field grows then exponentially owing to dynamo
action and saturates when the field reaches a certain
multiple of the equipartition value.
We solve the governing equations using the {\sc Pencil Code}%
\footnote{\texttt{http://pencil-code.googlecode.com/}}
which is a high-order finite-difference code (sixth order in space
and third order in time) for solving partial differential equations
on massively parallel machines.
Our model setup is identical to that used by \cite{KB09}, who studied
the frequency of dynamo waves in the saturated regime as a function of
the fractional helicity and thereby the effective dynamo number and,
more importantly, different magnetic field strengths.

\subsection{The VC flux in simulations}

We focus here on the results of four simulations, with parameters
given in \Tab{pars}.
Standard estimates suggest that the two dynamo parameters given in \Eq{dyncoef}
suffice for large-scale dynamo action ($C_\alpha C_S>2$), and also their
ratio, $C_S/C_\alpha\approx3 \Sh\ \kf/k_1$, is large enough for oscillatory
dynamo action with significant dynamo wave speeds --
even for Run D, with a ratio of unity (note the finite wavespeed $c=0.24\eta_{t0} k_z$). 
Recall that the vertical elongation of Run D suppresses non-vertical mean field structures
by increasing the minimum horizontal wavenumber to $4$, suppressing any
potential $x$-varying $\alpha^2$ field.

\begin{table}[b!]\caption{
Summary of the runs discussed in this paper.  The
wave speed $c$ is described in \Eq{wavespeed}.
}\vspace{12pt}\centerline{\begin{tabular}{ccccccccc}
Run & Fig. & $\Rm$ & $\Rey$ & $\Sh$ & $C_{\alpha} = \kf/k_1$
& $C_S$ & Resolution & $c/\eta_{t0}k_z$  \\
\hline
A & \ref{p_32f2} & $9$ & $0.4$ & $0.95$ & $5$ & 
                             $71$ &$64^3$ & $0.6$\\
B & \ref{p_shear64c} & $90$ & $ 9$ & $0.5$ & $3$ &
                             $14$ & $64^3$ & $1$\\
C & \ref{p_128kf3b} & $280$ & $19$ & $0.4$ & $3$ & 
                             $11$ &$128^3$ & $0.5$\\
D & \ref{p_256kf20c} & $7.9$ & $1.6$ & $0.016$ & $20$ &
                             $19$ & $64^2\times 256$ & $0.24$\\
\label{pars}\end{tabular}}\end{table}

The magnetic field is normalized to the equipartition value
\EQ
\Beq=(\mu_0\rho_0)^{1/2}\urms,
\EN
while, as suggested by \Eq{FVC} and \Sec{travelingwave}, $\hhf$ and $\FFfM$
are normalized to
\EQ
h_0 \equiv \kf^{-2}\Big(\bra{\meanB_x^2}\bra{\meanB_y^2}\Big)^{1/2},\quad
F_0 \equiv  \kf^{-2}S\bra{\meanBB^2}.
\EN
The brackets represent full volume averaging, here of already planar
averaged values.

\begin{figure}[t!]\begin{center}
\includegraphics[width=\linewidth]{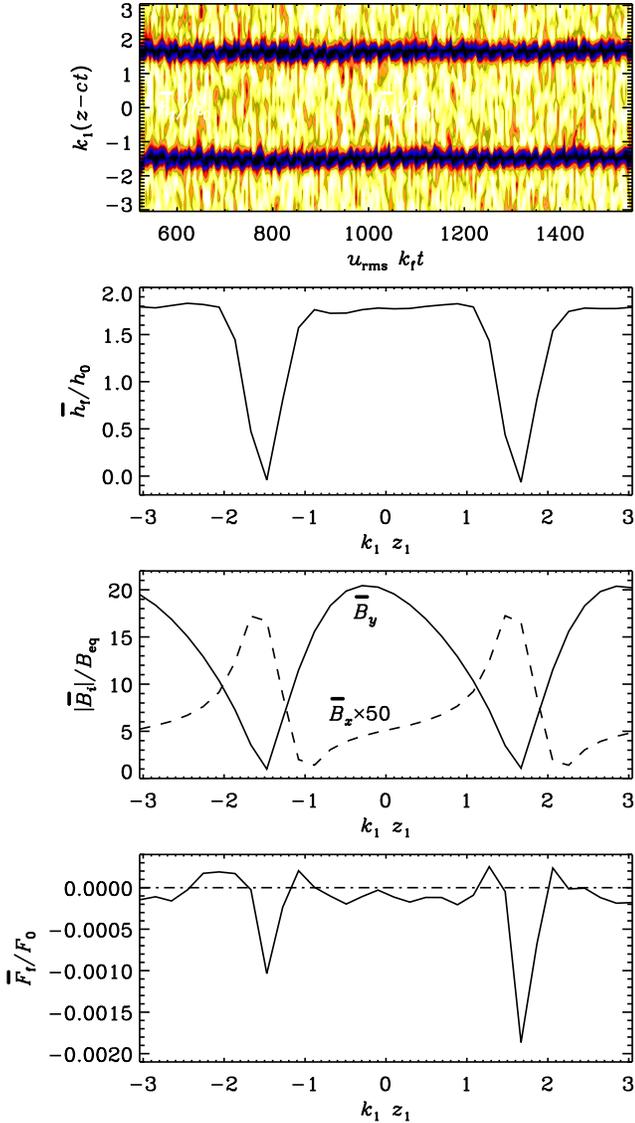}
\end{center}\caption{
Panel 1:
visualization of $\hf$ as a function of normalized $t$ and $z-ct$
for Run A: $\kf/k_1=5$, $\Rey=0.4$, $\Rm=9$, and $\Sh=0.95$.
Here $c=0.6\etatz k_1$ is the actual speed of the dynamo wave.
Panels 2--4 give the $z$ dependence of $\hf$, $\meanB_i$ for $i=x$, $y$,
and $\meanF$,
averaged in the comoving frame over the time of  the first panel.
}\label{p_32f2}\end{figure}

\begin{figure}[t!]\begin{center}
\includegraphics[width=\linewidth]{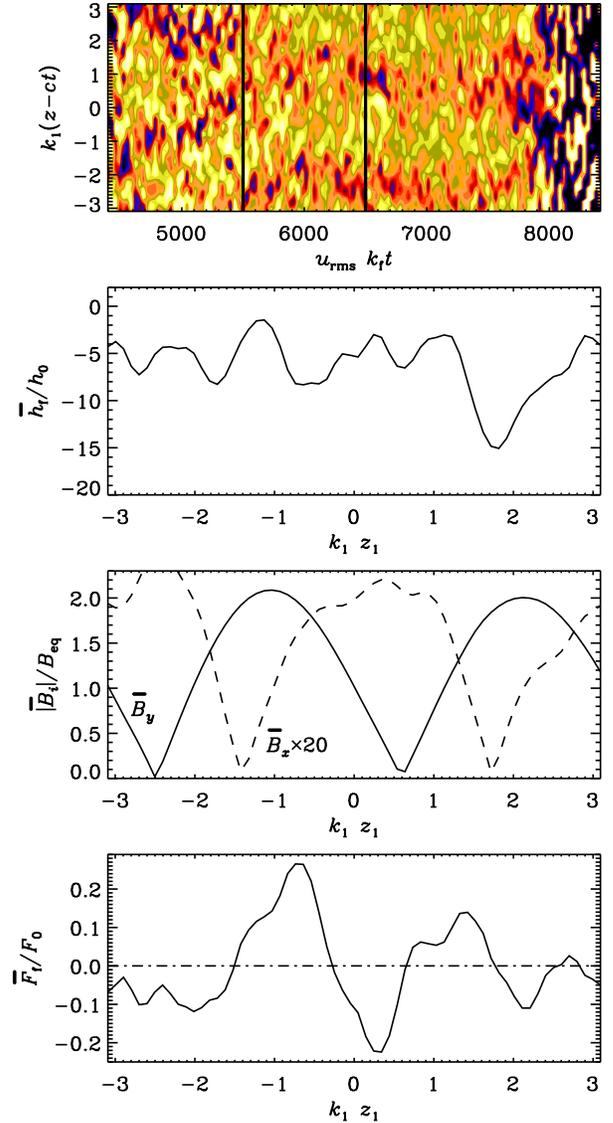}
\end{center}\caption{
Same as \Fig{p_shear64c}, but for 
Run B: $\kf/k_1=3$, $\Rey=9$, $\Rm=90$, and $\Sh=0.5$.
In this case, $c=1.04\etatz k_1$.
The vertical bars in the first panel denote the time interval over which
the functions in the other 3 panels are averaged.
Note also that after $\urms\kf t\approx7800$ the dominant dynamo mode
changes and the field becomes $x$ dependent.
}\label{p_shear64c}\end{figure}

\begin{figure}[t!]\begin{center}
\includegraphics[width=\linewidth]{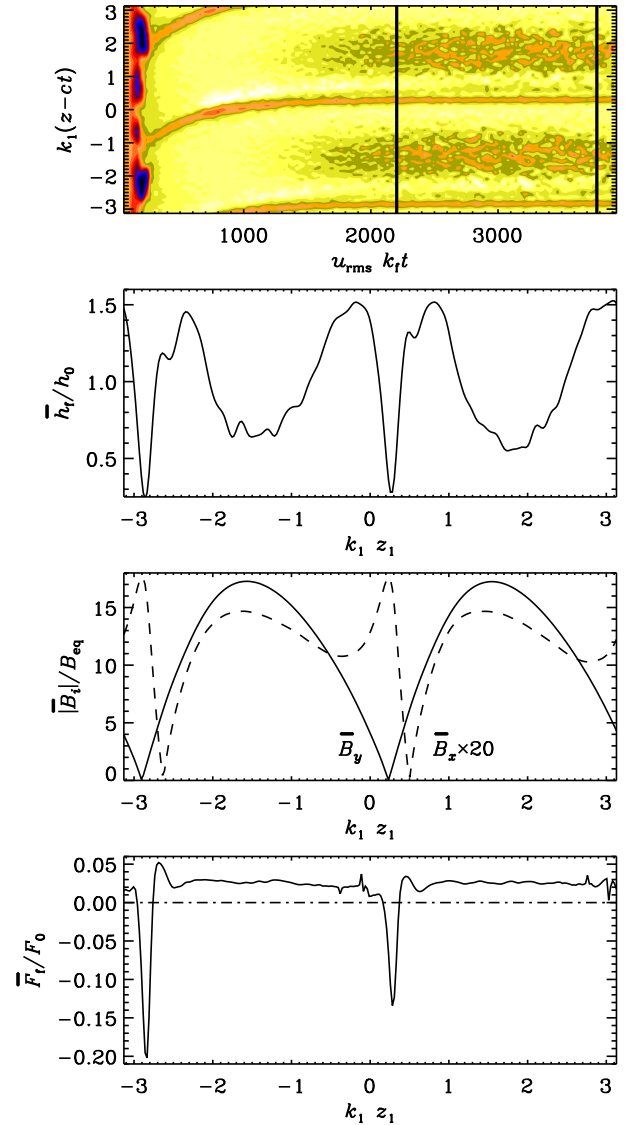}
\end{center}\caption{
Same as \Fig{p_shear64c}, but for 
Run D: $\kf/k_1=20$, $\Rey=9$, $\Rm=7.9$, and $\Sh=0.016$.
The domain for this run is $\pi/2\times \pi/2 \times 2\pi$,
and the helicity in the top panel has been multiplied by 10
to achieve dynamic color range.
The vertical bars in the first panel denote the time interval over which
the functions in the other 3 panels are averaged.
}\label{p_256kf20c}\end{figure}

The simulations developed the expected dynamo wave,
so we analyze the data in a comoving frame in which the
wave is standing, allowing us to average in time.
In agreement with earlier work \citep{Mitra10,HB10},
the magnetic helicity of the
small-scale magnetic field is then statistically steady and therefore the
divergence of the magnetic helicity flux must be independent
of the gauge chosen.
Note that for Run A (\Fig{p_32f2}) the flux is generally quite small (less than $10^{-3}$
times the value expected based on equation \ref{FVC}), except
when $\meanB_y=0$ where it shows a small peak.
The situation is different in Runs B and C (\Figs{p_shear64c}{p_128kf3b}),
where the flux is larger
but uncorrelated compared to the spatial dependence expected
to be dominated by $\meanB_y^2\gg\meanB_x^2$.
Finally note that for a mean-field wavenumber $k=k_1$ the expected
VC flux would be of wavenumber $2k_1$, not much smaller than forcing wavenumbers
of $3k_1$ or $5k_1$.  In \Fig{p_256kf20c} we therefore present the results of 
Run D with $\kf=20k_1$ in a non-cubic domain with $L_z=4L$.
The results are qualitatively
similar to Run A which has a similar $\Rm$.

\begin{figure}[t!]\begin{center}
\includegraphics[width=\linewidth]{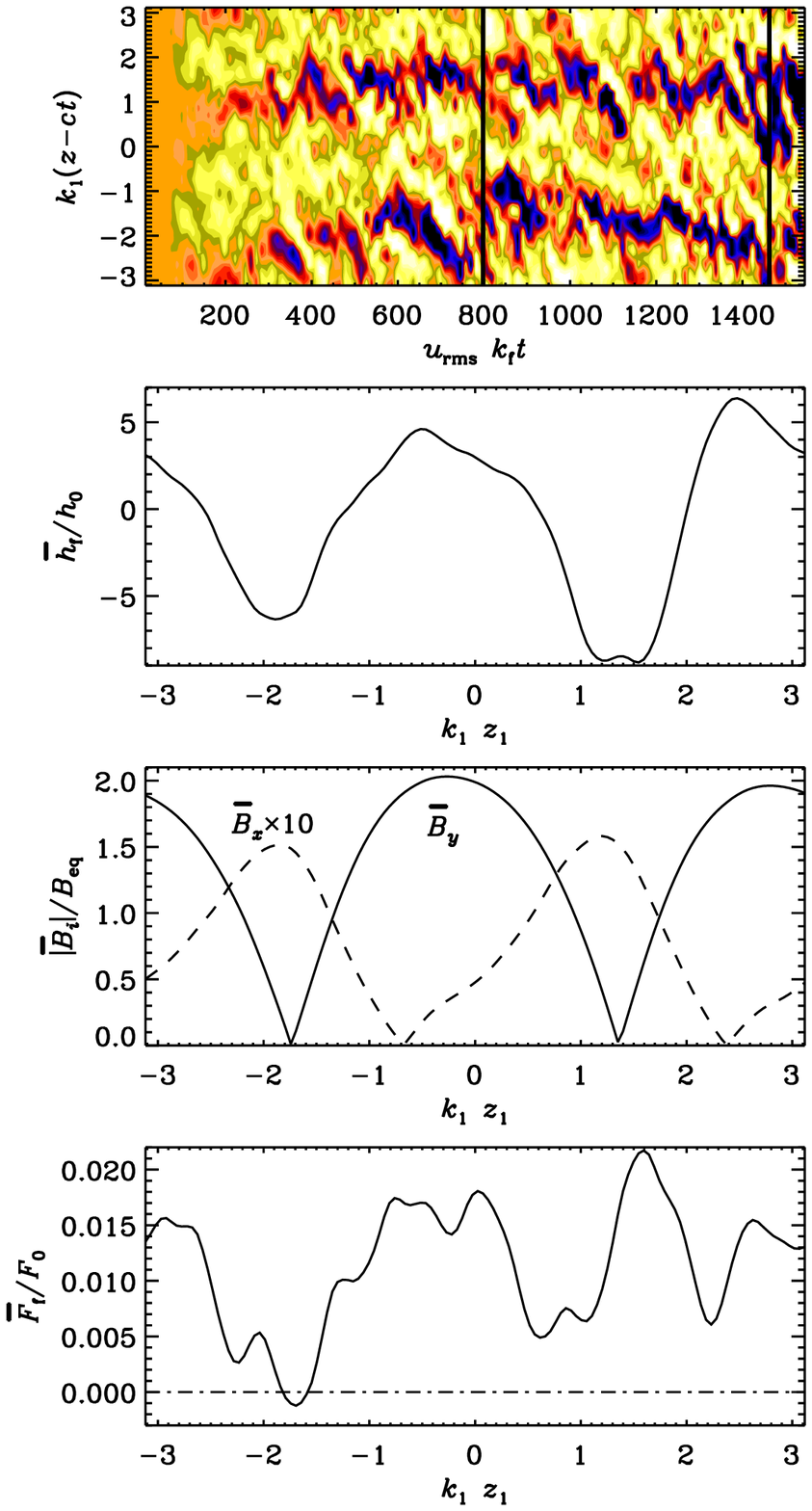}
\end{center}\caption{
Same as \Fig{p_shear64c}, but for 
Run C: $\kf/k_1=3$, $\Rey=19$, $\Rm=280$, and $\Sh=0.4$.
The vertical bars in the first panel denote the time interval over which
the functions in the other 3 panels are averaged.
}\label{p_128kf3b}\end{figure}

The significant result to draw from the figures is that the small-scale flux is both
smaller than the expected $\FVC$ and uncorrelated with it.  Note that as the
$y$-directed field is much greater than the $x$-directed field, as expected,
and that the
VC flux is approximately proportional to the square of the $\overline{B}_y$.
Thus, according to our present results we must conclude that there is no
support for the validity of \Eq{FVC}.

\section{Comparison with previous work}
 \label{comparison}

We would like to emphasize that our results are not in contradiction with previous calculations: we
are not working in a gauge where one would expect the VC flux to exist.
Disentangling the differing components (there are four in \citealt{SB06},
including an unexplored triple correlator) is not straightforward.
In this work we are using the gauge independence of the divergence of the
flux of small-scale magnetic helicity, as described in \Sec{significance},
to relate the observations in the shearing-advective gauge to the expected
divergence of $\FVC$.

To date, the only numerical evidence for the VC flux comes from
interpretations of the differing dynamo behavior
in shearing systems with vertical field boundary conditions (that allow a flux) as compared with those systems with
perfect conductor boundary conditions that disallow a flux; see \Eq{Fp}.
Examples of such indirect evidence include the papers by \cite{B05} and
\cite{KKB08}.
An alternative interpretation might simply be that the excitation condition
for the onset of large-scale dynamo action are simply delayed sufficiently
when changing the boundary condition from a vertical field to a
perfect conductor condition, as was discussed also by \cite{KKB10}.
It should also be noted that the use of $\FVC$ in various dynamical quenching
models has not alleviated catastrophic quenching unless $\CVC$ is
increased beyond a certain limit where the flux divergence leads to
a magnetic $\alpha$ effect that is more important than the kinematic
$\alpha$ effect \citep{BS05a,GCB10}; see also \App{MeanField}.

 \section{Conclusions}
 \label{conclusion}
 
We conclude that there is at present no evidence for a shear-driven
vertical flux of small-scale magnetic helicity in a gauge
where the only significant flux must be vertical.
We speculate that the finite divergence of the VC flux found
earlier in analytic studies using Coulomb and related gauges might be
a consequence of the gauge choice, which can generate unexpected horizontal
helicity fluxes that are not normally considered.
When the gauge choice is such that those horizontal fluxes
are transformed out, there is no remaining vertical flux.
It appears therefore that the VC flux does either not operate,
or it at least does not follow the expected functional form.
We note that diffusive fluxes have been found to exist, so
there do remain mechanisms that can export small-scale magnetic
helicity from a dynamo.

The simultaneous export of large- and small-scale helicity
at some relative level is inevitable \citep{BB03} and in fact necessary, because
otherwise simple considerations \citep{BDS02,BS05} have long suggested
that the magnetic energy would reach unrealistically large values.
The VC flux was a particularly promising mechanism to export
small-scale magnetic helicity
because it could do so while allowing the system to retain much of
the large-scale dynamo generated field, resulting in rapid growth to
strong mean fields.
Turbulent diffusion of magnetic helicity is
now the most promising escape from catastrophic $\alpha$ quenching --
even for shearing systems.
Simulation and theory suggest that this
will be significant starting near $\Rm=10^4$ \citep{Mitra10,HB10},
which, while astrophysically significant
eludes numerical verification at present.
However, diffusive fluxes must inevitably export scales of helicity at comparable
fractional rates, reducing expected final field strengths below those that might
have been hoped for with the VC flux.

\section*{Acknowledgements}

The National Supercomputer Center in Link\"oping and the Center for
Parallel Computers at the Royal Institute of Technology in Sweden
are acknowledged.
This work was supported in part by the Swedish Research Council,
grant 621-2007-4064, and the European Research Council under the
AstroDyn Research Project 227952.


\end{document}